\documentstyle[multicol,aps,epsfig]{revtex} % for multicolumn
\newcommand{\beq}{\begin{equation}}
\newcommand{\eeq}{\end{equation}}
\newcommand{\beqa}{\begin{eqnarray}}
\newcommand{\eeqa}{\end{eqnarray}}
\input epsf

\tighten       %Gives single-space (RevTex 3.0)

\begin{document}
\draft %prints PACS numbers in 
%\usepackage{epsfig}
%\usepackage{fancyhdr}
%\usepackage{fancybox}
%\usepackage{texdraw}
%\documentclass[12pt,a4paper]{article}
%\usepackage{latexsym,epsfig,graphicx}
%\makeatletter
%\edef\@scale{\ifcase\@ptsize\@m\or\magstephalf\or\magstep1\or\magstep2\fi}
%\font\msy=msbm10 scaled \@scale \newfam\msbm \textfont\msbm=\msy
%\def\bbold{\fam\msbm \msy} \makeatother

\title{Wavefronts may move upstream in semiconductor superlattices}

\author{A.~Carpio}
\address{Departamento de Matem\'atica
Aplicada, Universidad Complutense, Madrid 28040, Spain.}
\author{L.L.~Bonilla}
\address{Departamento de Matem\'aticas,   
Universidad Carlos III de Madrid,
Avenida de la Universidad 30, 28911 Legan{\'e}s, Spain}
\author{A.~Wacker and E.~Sch\"oll}
\address{Institut f\"ur Theoretische Physik, Technische Universit\"at Berlin,
Hardenbergstr.\ 36, 10623 Berlin, Germany}
\date{%November 29, 1999
\today 
}

\maketitle

\begin{abstract}
In weakly coupled, current biased, doped semiconductor superlattices, domain 
walls may move upstream against the flow of electrons. For appropriate doping 
values, a domain wall separating two electric field domains moves downstream 
below a first critical current, it remains stationary between this value and 
a second critical current, and it moves upstream above. These conclusions 
are reached by using a comparison principle to analyze a discrete 
drift-diffusion model, and validated by numerical simulations. Possible 
experimental realizations are suggested. 
\end{abstract}
\pacs{PACS numbers: 05.45.-a, 72.20Ht, 73.61.-r }

\begin{multicols}{2}
\narrowtext

\setcounter{equation}{0}
\section{Introduction}
\label{sec:intro}
Current instabilities in doped semiconductor superlattices (SL) have been 
an active subject of research during this decade. For strongly coupled SL,
Bloch oscillations \cite{esa70,but77,sib89} and Wannier-Stark hopping 
\cite{tsu75} produce negative differential conductivity (NDC) at high 
electric fields. This may result in self-sustained oscillations of the 
current due to recycling of charge dipole domains as in the Gunn effect 
of bulk n-GaAs \cite{gun63,but77}. For weakly coupled SL, sequential tunneling is 
the main mechanism of vertical transport. Under dc voltage bias conditions, 
stationary electric field domains may form if doping is large enough 
\cite{esa74,gra91}. Below a critical doping value, the existing charge inside 
the SL may not be able to pin domain walls, and current self-oscillations 
appear \cite{kas95,kas97}. These oscillations may be due to recycling of 
charge monopoles (domain walls) or dipoles depending on the boundary 
condition at the injecting contact region (in a typical n$^+$-n-n$^+$ 
configuration with the SL imbedded between highly doped regions, the 
doping at the emitter region is crucial) \cite{san99}. Driven chaotic 
oscillations have also been predicted \cite{bul95} and observed in 
experiments \cite{zha96}. Lastly, there are ways to tune the charge 
inside the SL (and therefore obtain stationary domains or self-oscillations) 
without replacing it by a different one. For example, by applying a transverse 
magnetic field \cite{sun99} or by photoexciting the SL \cite{oht98}. 

Transport in weakly coupled SL can be described by simple rate 
equation models for electron densities and average fields in the wells,
\cite{pren94,bon94,bon95,wac97a}. Many of the effects related above 
have been explained by means of a simple discrete drift model 
\cite{bon94,bon95,wac97,pat98}. In this model, the tunneling current between 
two adjacent wells, $J_{i\to i+1}$, equals the 2D electron charge density 
at well $i$ times a drift velocity, which depends on the electric field 
at the same well. By starting from a microscopic sequential tunneling model, 
it has been shown that the discrete drift model is a good approximation at low 
temperatures and for fields above the first plateau of the SL current-voltage
characteristic \cite{wac97a,bon99}. For low dc voltages on the first plateau, 
a discrete diffusion (which is a nonlinear function of the field) should be 
added. This term contains the contribution to $J_{i\to i+1}$ of the  
tunneling from well $i+1$ back to well $i$ (which vanishes for large enough
electric fields) \cite{wac97a,bon99}. In this paper we report an interesting 
consequence of electron diffusivity at low fields: if the current is 
sufficiently high, and so is the doping, a domain wall (monopole wave) 
which connects two domains may travel in a direction opposite to the flow 
direction for electrons (i.e., {\em upstream}, in the positive current 
direction!). This striking phenomenon is contrary to the usual situation: 
a monopole either moves downstream (in the direction of the flow of 
electrons), or it remains stationary, \cite{wac97}. 
We substantiate our claim both by numerical simulations of the discrete 
drift-diffusion model and by rigorous mathematical analysis based upon a 
comparison principle \cite{aro78}. Mathematical analysis yields useful 
bounds for critical values of current and well doping, and for monopole 
velocity. 

There are related fields for which differential-difference equations 
(similar to discrete drift-diffusion models) model the systems of 
interest. Well-known are propagation of nerve impulses along myelinated 
fibers, modelled by discrete FitzHugh-Nagumo equations \cite{kee98,keener}, 
motion of dislocations \cite{nab67,car99} and sliding charge density waves 
\cite{cdw}, modelled by variants of the Frenkel-Kontorova model \cite{FK},
etc. The theory of wavefront propagation has been developed for some of 
these models, which are simpler than ours: convection is typically 
absent from them and diffusion is purely linear \cite{keener}.

The rest of the paper is as follows. We write the drift-diffusion model 
with appropriate boundary conditions in Section \ref{sec:model}. There 
we render these equations dimensionless and explain the results of numerical 
simulations on a current biased infinitely long SL. Furthermore, we find 
by numerical simulations that our results for infinite SL may be realized 
in finite SL with appropriate boundary conditions under constant current 
bias. The theoretical analysis based on the comparison principle is 
presented in Section \ref{sec:math}. Section \ref{sec:dis} contains our 
conclusions. Finally some material of a more technical nature is
relegated to the Appendices. 

\section{Discrete drift-diffusion model}
\label{sec:model}
\subsection{Equations and boundary conditions}
At low enough temperatures (much less than a typical Fermi energy of a SL
well measured from the first subband, say 20 meV or 232 K), the following 
discrete drift-diffusion equations model sequential vertical transport in a 
weakly doped SL \cite{wac97a,bon99}:
\begin{eqnarray}
{\varepsilon\over e}\, {dF_{i}\over dt} + {n_{i}v(F_{i})\over d+w}
- D(F_i)\, {n_{i+1}-n_{i}\over (d+w)^{2}}= J(t)\,, \label{m1}\\
F_{i}-F_{i-1} = {e\over\varepsilon}\, (n_{i}-N_{D}^{w}) . \label{m2}
\end{eqnarray}
Eq.\ (\ref{m1}) is Amp\`ere's law establishig that the total current 
density, $e J$, is sum of displacement and tunneling currents. The latter
consists of a drift term, $e n_i v(F_i)/(d+w)$, and a diffusion term, 
$e D(F_i)\, (n_{i+1}-n_i)/(d+w)^2$. We have adopted the convention (usual 
in this field) that the current density has the same direction as the flow 
of electrons. Eq.\ (\ref{m1}) holds for $i=1,\ldots,N-1$. Eq.\ (\ref{m2}) 
is the Poisson equation, and it holds for $i=1,\ldots,N$. $n_i$ is the 2D 
electron number density at well $i$, which is singularly concentrated on a 
plane located at the end of the well. $F_i$ is {\em minus} an average electric 
field on a SL period comprising the $i$th well and the $i$th barrier (well $i$ 
lies between barriers $i-1$ and $i$: barriers 0 and $N$ separate the SL from 
the emitter and collector contact regions, respectively). Parameters 
$\varepsilon$, $d$, $w$, and $N_{D}^{w}$ are well permittivity, barrier 
width, well width and 2D doping in the wells, respectively. 

\begin{figure}
\begin{center}
%\centerline{\hbox{\psfig{figure=vd.ps,width=7cm}}}
\fbox{
\epsfig{file=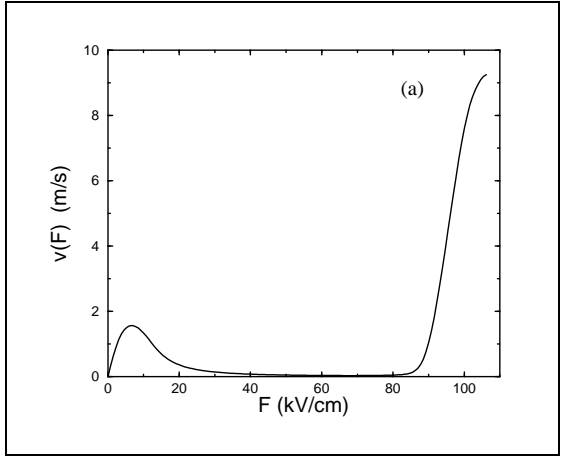, width=7cm}}
\vspace{0.5 cm}
\fbox{
\epsfig{file=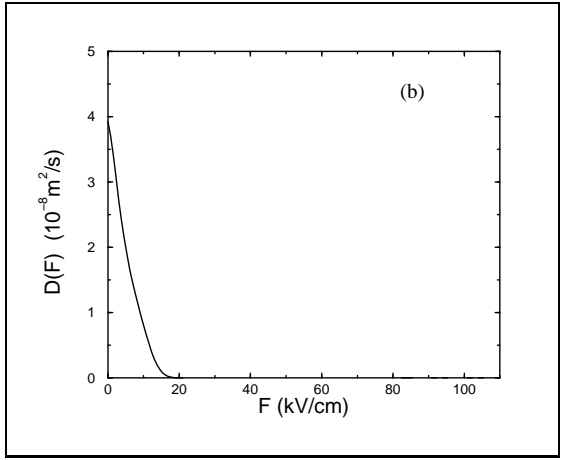, width=7cm}}
\vspace{0.5 cm}
\fbox{
\epsfig{file=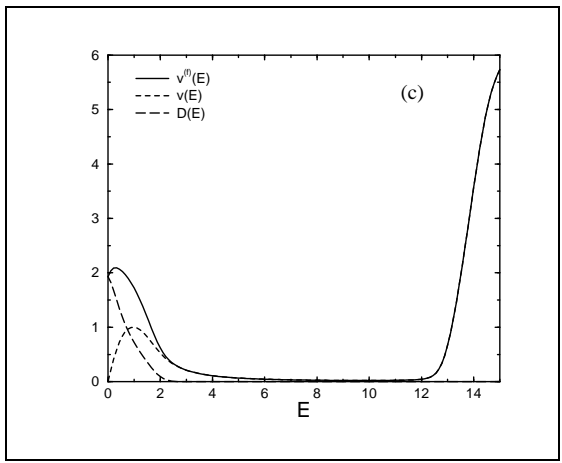, width=7cm}}
\vspace{0.5 cm}
\fbox{
\epsfig{file=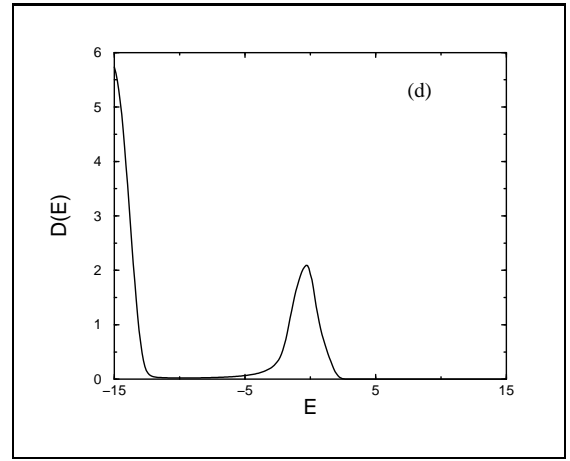, width=7cm}}
\vspace{0.5 cm}
\caption{(a) Drift velocity and (b) diffusion coefficient for a 
9nmGaAs/4nmAlAs SL. Doping at the wells is $N_{D}^{w} = 1.5\times 10^{11}$ 
cm$^{-2}$, whereas at the contact regions, $N_D  = 2\times 10^{18}$ 
cm$^{-3}$. (c) Dimensionless drift velocity, $v(E)$, diffusion (equivalent 
to backward tunneling velocity), $D(E)$, and forward tunneling velocity, 
$v^{(f)}(E) = v(E) + D(E)$. (d) Extension of the dimensionless diffusivity 
to negative values of field. We have $D(-E) = v^{(f)}(E)$. The same formula 
yields the extension of $v^{(f)}(E)$ to negative fields. Then $v(E)$ is an 
odd function of $E$.}
\label{f1}
\end{center}
\end{figure}

Drift velocity and diffusion coefficient are depicted in Fig.\ 
\ref{f1} for the 9nmGaAs/4nmAlAs SL of Ref.\ \onlinecite{kas97}. We have 
obtained them from microscopic calculations presented in Ref.\ 
\onlinecite{wac97a} (which is appropriate for these sample parameters 
\cite{wac98}) by setting $v(F) = J(N_D^w,N_D^w,F)\, (d+w)/N_D^w$
and $D(F)=-(\partial J(N_D^w,N_D^w,F)/\partial n_{i+1}) (d+w)^2$. Here 
$e\,J(n_i,n_{i+1},F_i)$ is the tunneling current between wells $i$ and $i+1$, 
$J_{i\to i+1}$. We assume that the tunneling current is a function of the 
average field at the $i$th SL period, $F_i=F$, and of the 2D electron densities 
at wells $i$ and $i+1$, $n_i$ and $n_{i+1}$, respectively. Notice that our 
model for the tunneling current,
\begin{eqnarray}
e J(n_i,n_{i+1},F_i) = {e n_i v(F_i)\over d+w} - e D(F_i)\, {n_{i+1} - n_{i} 
\over (d+w)^{2}}
\nonumber\\
\equiv {e n_i v^{(f)}(F_i) - e n_{i+1} v^{(b)}(F_i)\over d+w}\,,
\label{velocities}
\end{eqnarray}
is reasonable for temperatures much lower than a typical Fermi energy in the 
wells measured from the first subband (say 20 meV), \cite{bon99}. The tunneling 
current density should change sign if we reverse the electric field and 
exchange the electron densities at wells $i$ and $i+1$: $J(n_i,n_{i+1},F_i) 
= - J(n_{i+1},n_i,-F_i)$. This inversion symmetry implies 
$$v^{(f)}(-F) = v^{(b)}(F)\quad \mbox{and}\quad v(-F) = - v(F),$$ 
where $v^{(b)}(F) = D(F)/(d+w)$ and $v^{(f)}(F) = v(F) + v^{(b)}(F)$. See 
Figure \ref{f1}(d).

Equations (\ref{m1}) and (\ref{m2}) should be supplemented with appropriate 
bias, initial and boundary conditions. Among possible bias conditions, we 
shall consider the extreme cases of current bias ($J(t)$ specified) and 
voltage bias:
\begin{eqnarray}
(d+w)\, \sum_{i=1}^{N} F_i = V\,,\label{m3}
\end{eqnarray}
with specified $V=V(t)$. Using (\ref{m3}) ignores potential drops at the 
contact regions and at barrier 0, and it overestimates the contribution 
of barrier $N$ by a factor $1+w/d$ \cite{bon99}. These contributions are 
negligible for long SL ($N=40$ or larger), so that we shall adopt the 
simpler expression (\ref{m3}). Appropriate boundary conditions have been 
derived under the same approximations as in (\ref{m1}) \cite{bon99}. They 
are
\begin{eqnarray}
{\varepsilon\over e}\, {dF_{0}\over dt} + j_e^{(f)}(F_0) - {n_{1} 
w^{(b)}(F_{0})\over d+w} = J(t)\,, \label{m4}\\
{\varepsilon\over e}\, {dF_{N}\over dt} + {n_{N} w^{(f)}(F_{N})\over 
d+w} = J(t)\,, \label{m5}
\end{eqnarray}
where the emitter current density, $e\, j_e^{(f)}(F)$, the emitter 
backward velocity, $w^{(b)}(F)$, and the collector forward velocity, 
$w^{(f)}(F)$ are functions of the electric field depicted in Fig.\ 
3 of Ref.\ \onlinecite{bon99} for contact regions similar to those 
used in experiments \cite{kas97}. 

\begin{figure}
\begin{center}
\fbox{
\epsfig{file=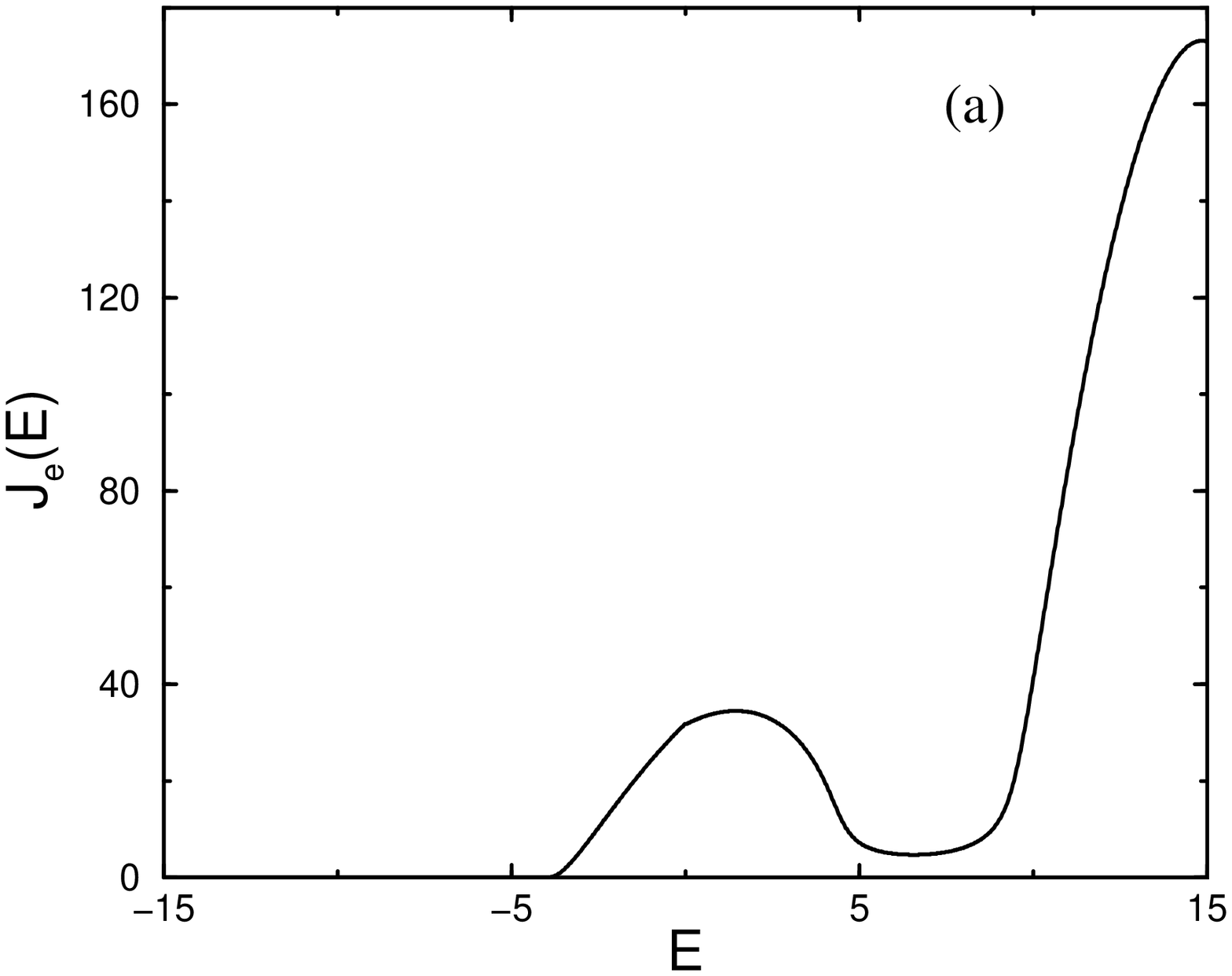, width=7cm}}
%\centerline{
%\epsfig{file=je.eps,angle=270,width=0.45\textwidth}
\vspace{0.5 cm}
%\centerline{
\fbox{
\epsfig{file=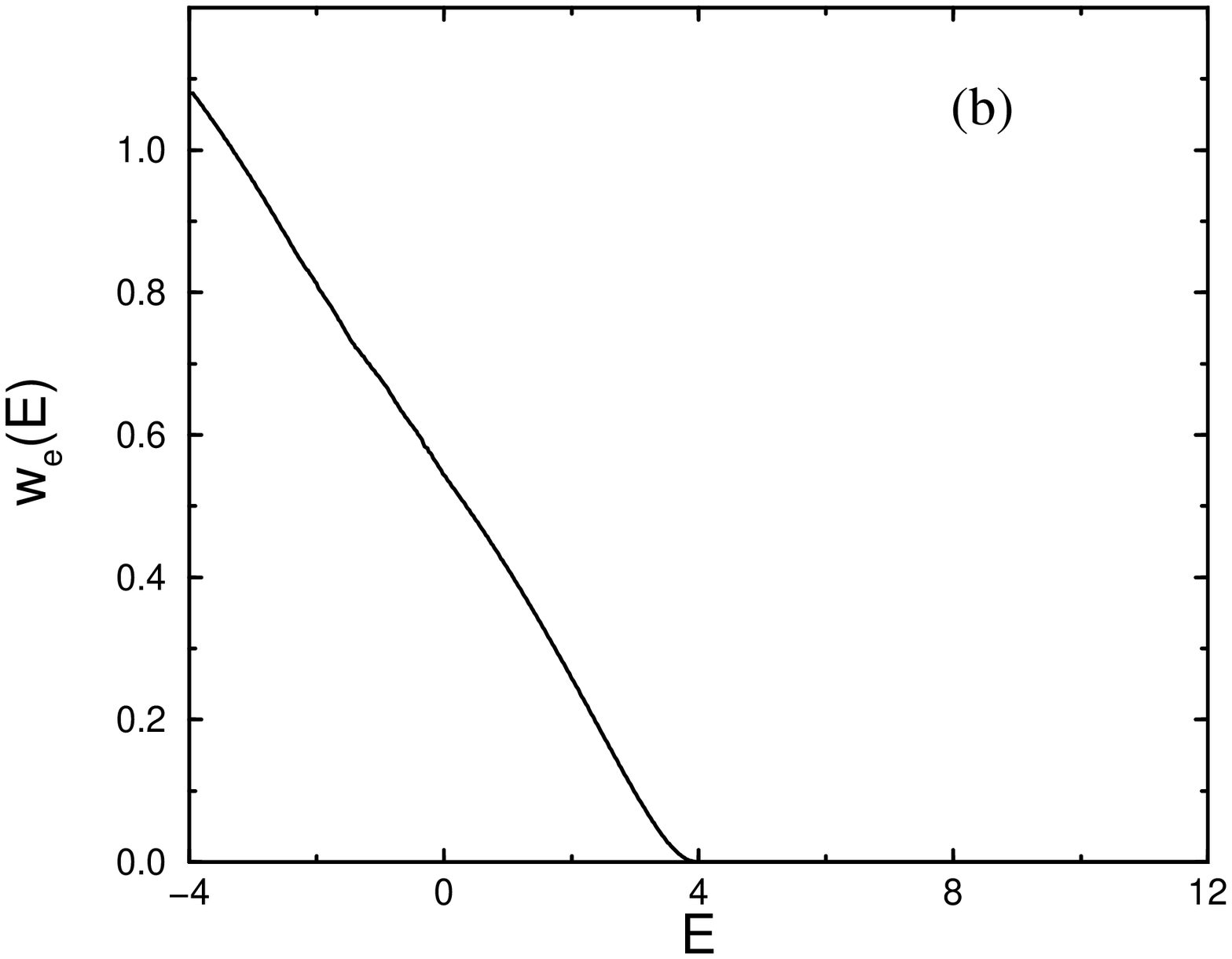,width=7cm}}
\vspace{0.5 cm}
%\centerline{
\fbox{
\epsfig{file=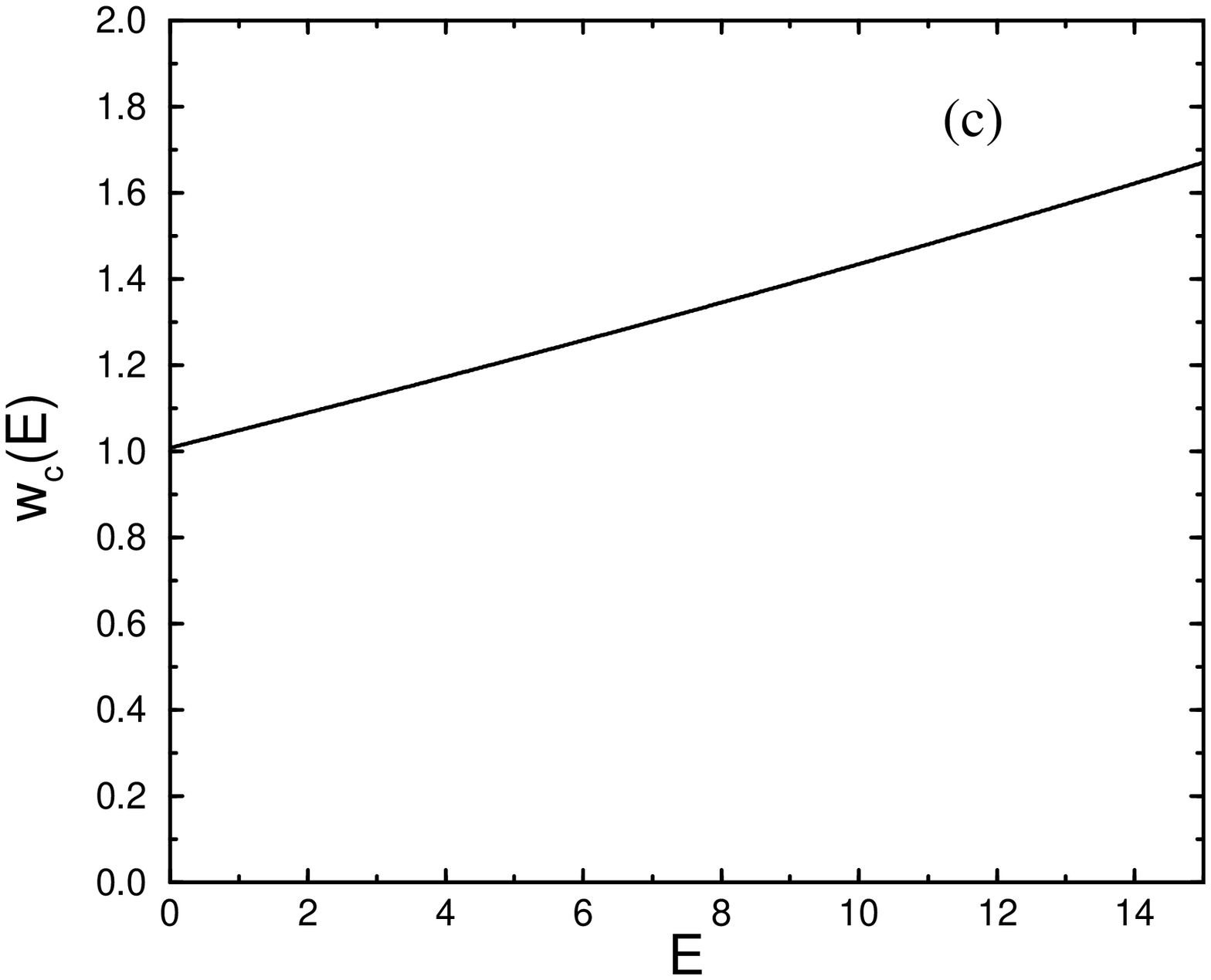,width=7cm}}
\vspace{0.5 cm}
%
%\centerline{\hbox{\psfig{figure=je.ps,width=7.0cm}}
%\hbox{\psfig{figure=je.ps,width=7.0cm}}}
%\centerline{\hbox{\psfig{figure=wb.ps,width=7.0cm}}}
%\hbox{\psfig{figure=wb.ps,width=7.0cm}}}
%\centerline{\hbox{\psfig{figure=wf.ps,width=7.0cm}}}
%
\caption{Dimensionless functions of the electric field for the contact 
regions. (a) Current at the emitter, $J_e(E)$. (b) Backward velocity at 
the emitter, $w_e(E)$. (c) Forward velocity at the collector, $w_c(E)$. }
\label{f2}
\end{center}
\end{figure}

To analyze the discrete drift-diffusion model, it is convenient to render 
all equations dimensionless. Let $v(F)$ reach its first positive maximum  
at $(F_M,v_M)$. We adopt $F_M$, $N_D^w$, $v_M$, $v_M\, (d+w)$, 
$eN_D^w v_M/(d+w)$ and $\varepsilon F_M (d+w)/(e N_D^w v_M)$ as the 
units of $F_i$, $n_i$, $v(F)$, $D(F)$, $eJ$ and $t$, respectively. 
For the first plateau of the 9/4 SL of Ref.\ \onlinecite{kas97}, we 
find $F_M = 6.92$ kV/cm, $N_D^w = 1.5\times 10^{11}$ cm$^{-2}$, $v_M = 
156$ cm/s, $v_M\, (d+w) = 2.03\times 10^{-4}$ cm$^2 /$s and $eN_D^w 
v_M/(d+w) = 2.88$ A/cm$^2$. The units of current and time are 0.326 mA 
and 2.76 ns, respectively. Then (\ref{m1}) to (\ref{m3}) become 
\begin{eqnarray}
{dE_{i}\over dt} + v(E_i)\, n_i - D(E_i)\, (n_{i+1}-n_i) = J, 
\label{e1}\\
E_i - E_{i-1} = \nu\, (n_i - 1) , \label{e2}\\
{1\over N}\,\sum_{i=1}^N E_i = \phi. \label{e3}
\end{eqnarray}
Here we have used the same symbols for dimensional and dimensionless 
quantities except for the electric field ($F$ dimensional, $E$ dimensionless). 
The parameters $\nu=e N_D^w/(\varepsilon\, F_M)$ and $\phi = V/[F_M N (d+w)]$ 
are dimensionless doping and average electric field (bias), respectively. For 
the 9/4 SL, $\nu\approx 3$. We recall that $i=1,\ldots,N-1$ in (\ref{e1}) 
and $i=1,\ldots,N$ in (\ref{e2}). The boundary conditions (\ref{m4}) and 
(\ref{m5}) become 
\begin{eqnarray}
{dE_{0}\over dt} + J_e(E_0) - w_e(E_0)\, n_1 = J, \label{e4}\\
{dE_{N}\over dt} + w_c(E_N)\, n_N = J, \label{e5}
\end{eqnarray}
where 
\begin{eqnarray}
J_e(E_0) = {j_{e}^{(f)}(F_{M}\, E_{0})\, (d+w)\over N_{D}^{w}	
v_{M}}\,, \nonumber\\ 
w_e(E_0) = {w^{(b)}(F_{M}\, E_{0})\over v_{M}}\,,
\nonumber\\
w_c(E_N) = {w^{(f)}(F_{M}\, E_{N})\over v_{M}}\,. \label{e6}
\end{eqnarray}
Figure \ref{f2} shows $J_e$, $w_e$ and $w_c$ as functions of the electric  
field. They are dimensionless versions of the curves plotted in Figure 3 
of Ref.\ \onlinecite{bon99}.

\subsection{Numerical simulations}
Simple solutions of the drift-diffusion equations (\ref{e1}) - (\ref{e2}) 
under constant current bias are stationary or moving monopole wavefronts 
connecting two electric field domains. Let us consider monopole solutions with 
profiles $\{E_i \}$ which are increasing functions of $i$, for they are 
compatible with realistic boundary conditions in which the emitter region
is highly doped \cite{kas97}. We have simulated numerically on a large SL,
\begin{eqnarray}
{dE_{i}\over dt} - {D(E_i) + v(E_i)\over \nu}\, (E_{i-1}-E_i) 
\nonumber\\
- {D(E_i)\over \nu}\, (E_{i+1}-E_i) = J-v(E_i), 
\label{e7}
\end{eqnarray}
with fixed $J$, which is equivalent to (\ref{e1}) - (\ref{e2}). Let 
$E^{(1)}(J)<E^{(2)}(J)<E^{(3)}(J)$, be the three solutions of $v(E)=J$ 
for $v_m < J< 1$, where $(E_m,v_m)$ is the minimum of $v(E)$ for $E>1$. 
For the 9/4 SL of Fig.\ \ref{f1}, $E_m=9.8571$, $v_m=0.02192$. We have 
simulated (\ref{e7}) for different values of $\nu>0$ and of $J\in 
(v_m,1)$. The initial condition was chosen 
so that $E_i \to E^{(1)}(J)$ as $i\to -\infty$, and $E_i \to E^{(3)}(J)$ 
as $i\to \infty$. We observed that, after a short transient, a variety of 
initial conditions sharing these features evolved towards either a 
stationary or moving monopole. For systematic numerical studies, 
we therefore adopted an initial step like profile, with $E_i=E^{(1)}(J)$ 
for $i<0$, $E_i=E^{(3)}(J)$ for $i>0$ and $E_0=E^{(2)}(J)$. The boundary data 
were taken to be $E_{-N}=E^{(1)}(J)$, $E_{N}=E^{(3)}(J)$ with $N$ large. 

Our results show that the dimensionless doping $\nu$ determines the type
of solution of (\ref{e7}) which is stable. There are two important 
values of $\nu$, $\nu_1<\nu_2$. 
\begin{itemize}
\item For $0<\nu<\nu_1$ and each fixed 
$J\in (v_m,1)$, only traveling monopole fronts moving downstream (to the right) 
were observed. For $\nu>\nu_1$, stationary monopoles were found. According to the 
arguments of Wacker et al \cite{wac97} for the discrete drift model with 
$D(E)=0$, stationary monopoles exist for dimensionless doping larger than a 
critical value. An upper bound for this critical doping is 
\begin{eqnarray}
\nu_c = v_m\, {E_{m}-1\over 1-v_{m}}\,, \label{m7}
\end{eqnarray}
which equals $\nu_c = 0.198$ for our numerical example. We have found that 
$\nu_1 = 0.16$. This agreement with results obtained assuming $D(E)=0$ is 
not surprising: we shall prove in Section \ref{sec:math} that (\ref{m7}) 
holds as well for the model (\ref{e1}) - (\ref{e2}) with nonzero diffusivity. 
\item For $\nu_1<\nu<\nu_2$, traveling fronts moving downstream exist 
only if $J\in (v_{m},J_1(\nu))$, where $J_1(\nu)<1$ is a critical value 
of the current. If $J \in (J_1(\nu),1)$, the stable solutions are 
steady fronts (stationary monopoles). We have found that $\nu_2=0.33$. 
\item New solutions are observed for $\nu>\nu_2$. As before, there 
are traveling fronts moving downstream if $J\in (v_{m},J_1(\nu))$, and 
stationary monopoles if $J \in (J_1(\nu),J_2(\nu))$, $J_2(\nu)<1$ is a new 
critical current. For $J_2(\nu)<J<1$, the stable solutions of (\ref{e7}) 
are monopoles traveling upstream (to the left). As $\nu$ increases,
$J_1(\nu)$ and $J_2(\nu)$ approach $v_m$ and 1, respectively. Thus stationary 
solutions are found for most values of $J$ if $\nu$ is large enough. 
\end{itemize}

Figure \ref{f3} depicts $J_1(\nu)$ and $J_2(\nu)$ as functions of $\nu$. 
Notice that $J_1$ decreases from $J_1=1$ to $J_1=v_m$ as $\nu$ increases 
from $\nu_1$. Similarly, $J_2$ decreases from $J_2=1$ to a minimum value 
$J_2\approx 0.53$ and then increases back to $J_2=1$ as $\nu$ increases. 
Monopole velocity as a function of current has been depicted in Figure 
\ref{f4} for four different doping values, $\nu=0.5$, $\nu=1$, $\nu = 3$ and 
$\nu=10$. For larger $\nu$, the interval of $J$ for which stationary solutions
exist becomes wider again, trying to span the whole interval $(v_m,1)$ as 
$\nu\to\infty$. For very large $\nu$, the velocities of downstream and 
upstream moving monopoles become extremely small in absolute value.

Notice that if we use the complete sequential tunneling current instead 
of the drift-diffusion approximation (\ref{velocities}) in Eq.\ (\ref{m1}), 
the situation is the same. Figure \ref{f5} depicts monopole velocity versus 
current for well doping corresponding to the 9/4 SL of Ref.\ 
\onlinecite{kas97}. Results obtained with the complete sequential tunneling 
current or with approximation (\ref{velocities}) (corresponding to Fig.\ 
\ref{f4} with $\nu=3$) are compared. Both velocity curves are similar, and 
their quantitative discrepancies are irrelevant in view of the uncertainties 
involved in a theoretical calculation of the tunneling current (typically 
the off-resonance current is larger than the theoretical prediction). 
%Of course, the 
%drawback of using the complete sequential tunneling current is that we would 
%not have a mathematical theory to help us understanding the data of numerical 
%simulations. 

Once different stable monopole solutions (moving either downstream or 
upstream, stationary) have been identified, we raise the natural question 
of whether they are compatible with boundary conditions. Another series 
of numerical simulations was carried out to answer this. We solved 
numerically (\ref{e7}) for a current biased finite SL ($N=40$) with boundary 
conditions (\ref{e4}) - (\ref{e6}). Our results are depicted in Figure 
\ref{f6} for realistic doping at the contact layers. We observe that 
the emitter boundary condition results in the creation of a charge 
accumulation layer near this contact. A charge depletion layer is formed 
near the collector contact as a result of the corresponding boundary 
condition. Except for these layers, existence and configuration of 
monopoles moving downstream, upstream or remaining stationary,
agrees with the previous simulations (corresponding to an infinitely 
long current-biased SL with a monopole-like initial condition). 

\section{Mathematical analysis of traveling monopoles and stationary solutions}
\label{sec:math}
In this Section, we study theoretically moving or stationary monopoles on 
an infinitely long, current-biased SL. Our findings will confirm the picture 
suggested by the numerical simulations of the previous Section for any 
doped weakly coupled SL. Furthermore, we shall prove stability of the 
different monopole solutions and find bounds for the critical values of 
$\nu$ and $J_i$. Our results are based upon and extend ideas first proposed 
by J.P.\ Keener for discrete FitzHugh-Nagumo equations, corresponding to 
signal transmision in myelinated neurons \cite{keener}. Mathematically 
analogous problems arise in models of propagation of defects in crystals
\cite{nab67}. These problems have the following structure,
\begin{eqnarray}
{dE_{i}\over dt} - d\, (E_{i+1}-2E_i+E_{i-1}) = J-v(E_i), \label{simple}
\end{eqnarray}
which is much simpler than (\ref{e7}). Here the parameter $d>0$ is 
a constant diffusion coefficient, and $v(E)$ a `cubic' function with 
three branches as the electron drift velocity of Fig.\ \ref{f1}. 

For (\ref{simple}), there are critical values of $J$, $J_1$ and $J_2$, 
characterizing wavefront behavior \cite{keener}. For $J>J_2(d)$, there 
exist wavefront solutions of (\ref{simple}) moving upstream (to the left). 
For $J<J_1(d)$, there are wavefronts moving downstream (to the right), 
whereas for $J_1(d)<J<J_2(d)$, stationary fronts exist. The width of the 
interval $(J_1(d),J_2(d))$ is an increasing function of $d$. 

\subsection{Propagation failure and stationary solutions}
In Appendix \ref{theorem} we state and prove a comparison principle for 
(\ref{e7}). As a consequence, if our initial field profile is monopole-like 
[monotone increasing with well index, and sandwiched between $E^{(1)}(J)$ 
and $E^{(3)}(J)$], so is the electric field profile for any later time $t>0$; 
see Appendix \ref{theorem}: 
\begin{eqnarray}
\{E_i(0)\} \quad\mbox{increasing with $i$}\quad \Longrightarrow \nonumber\\
E^{(1)}(J)<E_i(t)<E_{i+1}(t)<E^{(3)}(J),\,\forall i,\, t>0.\nonumber
\end{eqnarray}
We now obtain sufficient conditions for an initial monopole {\em not to 
propagate} upstream or downstream. Under these conditions, the monopole 
may remain stationary or move downstream or upstream, respectively. Let us 
start with a condition pinning the left tail of a monopole. As $E_{i-1}<E_i$ 
and $E_{i+1} < E^{(3)}(J)$, we have:
\begin{eqnarray}
{dE_{i}\over dt} &=& {D(E_{i})\over\nu}\, (E_{i+1}-E_{i})\nonumber\\
&+& {D(E_{i})+v(E_{i})\over\nu}\, (E_{i-1}-E_i) + J-v(E_i)\nonumber \\
&\leq & {D(E_{i})\over\nu}\, [E^{(3)}(J)-E_i] + J-v(E_i) \leq 0, \nonumber
\end{eqnarray}
provided there exist $a_l<b_l$ such that 
\begin{eqnarray}
{D(E)\over \nu}\, [E-E^{(3)}(J)]\geq J-v(E),\quad E \in(a_l,b_l),
\label{c-} 
\end{eqnarray}
and then we choose some initial field, $E_i(0)\in (a_l,b_l)$. The previous 
inequality then implies $E_i(t)\in (E^{(1)}(J),b_l)$ for all $t>0$. This 
in turn forbids a monopole to move upstream (to the left). We say that 
condition (\ref{c-}) {\em pins} the left tail of the monopole. Whether 
such $(a_l,b_l)$ exist, depends on the parameters $\nu$ and $J$; see 
Figure \ref{f7}.

Let us now pin the right tail of a monopole. As $E_{i+1}>E_i$ and $E_{i-1} > 
E^{(1)}(J)$, we have 
\begin{eqnarray}
{dE_{i}\over dt} &=& {D(E_{i})\over\nu}\, (E_{i+1}-E_{i})\nonumber\\
&+& {D(E_{i})+v(E_{i})\over\nu}\, (E_{i-1}-E_i) + J-v(E_i)\nonumber \\
&\geq & {D(E_{i})+v(E_{i})\over\nu}\, [E^{(1)}(J)-E_i] + J-v(E_i) \geq 0, 
\nonumber
\end{eqnarray}
provided there exist $a_r<b_r$ such that 
\begin{eqnarray}
{D(E)+v(E)\over \nu}\, [E-E^{(1)}(J)]\leq J-v(E),\nonumber\\ 
E \in (a_r,b_r),    \label{c+} 
\end{eqnarray}
and we choose some initial field, $E_i(0)\in (a_r,b_r)$. The previous 
inequality then implies $E_i(t)\in (a_r,E^{(3)}(J))$ for all $t>0$. 
A monopole cannot then move downstream (to the right), and we say 
that its right tail is pinned. Figure \ref{fpin} illustrates our arguments:
for fields larger than $a_r$, the $E_i$'s tend to increase above $a_r$ 
toward $E^{(3)}(J)$. Then the monopole cannot move downstream. For $E_i <
b_l$, the fields tend to $E^{(1)}(J)$, and the monopole cannot move 
upstream. As before, the existence of $(a_r,b_r)$
depends on the values of $\nu$ and $J$; see Figure \ref{f7}. 

Figures \ref{f7} show the curves $J-v(E)$, $D(E)\, (E-E^{(3)})/\nu$ 
and $[D(E)+v(E)]\, (E-E^{(1)})/\nu$ for $\nu=3$ and different values of $J$. 
At $J=0.08$, Figure \ref{f7}(a) shows that there is an interval $(a_l,b_l)$ 
as in (\ref{c-}), but no interval $(a_r,b_r)$ as in (\ref{c+}) exist. Then 
the left tail of a monopole is pinned, but its right tail is free. In this 
conditions, a monopole may move downstream. Figure \ref{f7}(b) shows a monopole
with both its left and right tails pinned for $J=0.2$. Then our theory 
implies that wavefront propagation {\em fails} and a monopole-like stationary 
solution is stable. Numerical simulations show that there are stationary 
solutions when $J\in [0.09,0.53]$. Finally, Figure \ref{f7}(c) shows that, 
if $J=0.34$, the right tail of a monopole is pinned, but not its left tail. 
Under these conditions a monopole may move upstream. For larger $\nu$, the 
estimates become sharper. For instance, when $\nu=10$, (\ref{c-}) and 
(\ref{c+}) hold for $J\in [0.05,0.45]$. Direct numerical simulations show 
that stationary solutions exist for $J\in [0.04,0.55]$. Systematic use of 
these criteria allows us to estimate the critical doping values $\nu_j$ and 
critical current values $J_i(\nu)$, $i=1,2$ defined in the previous Section; 
see Appendix \ref{nuc}. Instead of looking for $J_1(\nu)$ and $J_2(\nu)$, it 
is more convenient to look for their inverse functions, which we may call 
$\nu_1(J)$ and $\nu_2^{\pm}(J)$. According to Figure \ref{f3}, the inverse 
function of $J_2(\nu)$ is two-valued, and its two branches are $\nu_2^{-}(J)
<\nu_2^{+}(J)$. We have found the following upper bounds $\nu_{1b}(J)$ and 
$\nu_{2b}^{+}(J)$ for $\nu_1(J)$ and $\nu_2^{+}(J)$, respectively: 
\begin{eqnarray} 
\nu_{1b}(J) = v_m\, {E_{m}-E^{(1)}(J)\over J - v_{m}}\,,
\label{nu1}\\
\nu_{2b}^{+}(J) = D(1)\, {E^{(3)}(J) - 1\over 1 - J}\,.  \label{nu2}
\end{eqnarray} 
If $\nu>\nu_{1b}(J)$, the right tail of the monopole is pinned, whereas 
the left tail of the monopole is pinned if $\nu>\nu_{2b}^{+}(J)$; 
see Appendix \ref{nuc}. Notice that $\nu_{1b}(J)$ is a decreasing 
function of $J$. Therefore the critical value $\nu_1$ (above which there 
are stationary solutions) is smaller than $\nu_{1b}(1)$, {\em which is exactly 
Wacker et al's bound}, (\ref{m7}). This explains why the bound (\ref{m7}) 
gives surprisingly good results even for the first plateau of the SL 
current-voltage characteristics [despite having been obtained under the 
assumption $D(E)\equiv 0$] \cite{wac97,wac97a}. Notice that the bound 
(\ref{nu2}) is reasonable for large dopings and currents $J\approx 1$. 
See Figure \ref{f3} for a comparison between the critical curves $J_1(\nu)$ 
and $J_2(\nu)$ and the bounds (\ref{nu1}) and (\ref{nu2}). 

\subsection{Propagation: traveling fronts}
Having shown that only one tail of a monopole is pinned suggests that 
the monopole may move in the opposite direction. Direct simulations 
show that this is often the case, and we will prove this now. 

An upstream traveling wave solution of (\ref{e7}) may have the form
\begin{eqnarray} 
E_i(t) = w(i+ct), \; c>0. \label{e8}
\end{eqnarray}
We will look for an electric field profile $w(z)$, $z=i+ct$, which is
not an exact solution of (\ref{e7}), but instead it satisfies 
\begin{eqnarray}
c\, {dw\over dz}  - {D(w(z)) + v(w(z))\over \nu}\, [w(z-1)-w(z)] 
\nonumber\\
- {D(w(z))\over \nu}\, [w(z+1)-w(z)] + v(w(z)) - J\leq 0.  
\label{e9}
\end{eqnarray}
If this {\em subsolution} is initially below an initial field profile,
i.e.\ $w(i)< E_i(0)$, for all $i$, then the comparison theorem of 
Appendix \ref{theorem} guarantees that $E_i(t)> w(i+ct)$ for later times. As 
$w(i+ct)$ moves upstream, so does $E_i(t)$, and the electric field 
profile corresponds to a monopole moving upstream with velocity at least 
$c$. See Figure \ref{fsbsp}: a subsolution ``pushes'' the monopole upstream,
whereas a supersolution (defined below) ``pushes'' the monopole downstream.

How do we find a reasonable subsolution? An idea is to try a piecewise 
continuous solution which equals $E^{(1)}(J)$ for $z<z_0$ and a larger  
constant $A$, $A\in (E^{(2)}(J),E^{(3)}(J))$ [and therefore $v(A)-J\leq 0$], 
for $z>z_1$, with $z_1>z_0$. 
For $z_0<z<z_1$, $w(z)$ is an unspecified smooth increasing function with
$w(z_0) = E^{(1)}(J)$ and $w(z_1) = A$. Now we shall select conveniently 
the numbers $z_0$, $z_1$, $c$ and $A$ so that (\ref{e9}) holds. Clearly, 
(\ref{e9}) holds for $z+1<z_0$ and for $z-1>z_1$. Suppose that $0<z_1 - 
z_0 < 1$. Then there are five possibilities:
\begin{enumerate}
\item $z<z_0$, $z+1>z_1$. Then $w(z-1)=w(z)=E^{(1)}(J)$, $w(z+1)=A$, which 
inserted in (\ref{e9}) yields $-D(E^{(1)})\, (A-E^{(1)})/\nu\leq 0$ 
(obviously true).
\item $z-1<z_0$, $z>z_1$. Then $w(z-1)=E^{(1)}(J)$, $w(z)=w(z+1)=A$, which 
inserted in (\ref{e9}) yields 
\begin{eqnarray}
J-v(A) \geq {D(A)+v(A)\over \nu} [A-E^{(1)}(J)]. \label{e10}
\end{eqnarray}
\item $z_0<z-1<z_1$. Then $E^{(1)}(J)<w(z-1)<A$ and $w(z)=w(z+1)=A$, which 
yields $J-v(A) \geq [D(A)+v(A)]\, [A-w(z-1)]/\nu$. This inequality holds 
if (\ref{e10}) does. 
\item $z_0<z<z_1$. Then $w(z-1) = E^{(1)}(J)$, $E^{(1)}(J)<w(z)<A$ and 
$w(z+1)=A$. Inserting this in (\ref{e9}), we find 
\begin{eqnarray}
c\, {dw\over dz}  \leq {D(w(z)) + v(w(z))\over \nu}\, [E^{(1)}-w(z)] 
\nonumber\\
+ {D(w(z))\over \nu}\, [A-w(z)] + J - v(w(z)) . \nonumber 
\end{eqnarray}
Let us now assume that we can select $A\in (E^{(2)},E^{(3)})$ such that
the right hand side of this expression is positive, say
\begin{eqnarray}
{D(w) + v(w)\over \nu}\, [E^{(1)}(J)-w] 
+ {D(w)\, (A-w)\over \nu} \nonumber\\
+ J - v(w)\geq \delta>0 ,\, E^{(1)}<w<A, \label{e11}
\end{eqnarray}
and that we choose $c$ so that $c\, dw/dz < \delta$. Then (\ref{e9}) holds. 
\item $z_0<z+1<z_1$. Then $w(z-1) = w(z) = E^{(1)}(J)$, and $E^{(1)}(J)<
w(z+1)<A$, which inserted in (\ref{e9}) yields $-D(E^{(1)})\, [w(z+1)-
E^{(1)}]/\nu\leq 0$ (obviously true). 
\end{enumerate}
Summarizing the previous arguments, provided (\ref{e10}) and (\ref{e11}) 
hold, $w(z)$ is a subsolution obeying (\ref{e9}). 
The parameter $A$ can be found graphically. First of all, we depict the 
functions $J-v(E)$ and $f_1(E;J)\equiv [D(E)+v(E)]\, [E-E^{(1)}(J)]/\nu$. 
Possible values of $A$ are those $E$ for which $J-v(E)\geq f_1(E;J)$. 
For such A, we may plot the left side of (\ref{e11}),
\begin{eqnarray}
f_2(E;J,A)\equiv {D(E) + v(E)\over \nu}\, [E^{(1)}(J)-E] \nonumber\\
+ {D(E)\, (A-E)\over \nu} + J - v(E).\label{e12}
\end{eqnarray}
If $f_2(E;J,A)>0$ for $E\in (E^{(1)}(J),A)$, then the selected value of $A$ 
allows us to construct the sought subsolution. See Figure \ref{f8} for a 
practical realization of this graphical construction. 

We have proved rigorously that monopoles may move upstream under favorable 
circumstances. Our proof using subsolutions may yield a very practical 
additional bonus: an upper bound, $c^{*}$, for the velocity of the monopole. 
Let us choose $\delta(J,A) = $ min$_{E^{(1)}<E<A} f_2(E;J,A)$, $z_1-z_0=1$, 
and $w(z)=[A-E^{(1)}(J)] (z-z_0)$ for $z_0<z<z_1$. Then $c^{*} = 
\delta(J,A)/[A-E^{(1)}(J)]$. In Figure \ref{f4}, $-c^{*}$ is represented 
by a line of thick dots for doping $\nu=3$ corresponding to the 9/4 SL.

In a similar vein, we can construct {\em supersolutions} which push the 
monopole field profile to the right; see Figure \ref{fsbsp}. Now we 
start from a monopole profile moving downstream, 
\begin{eqnarray} 
E_i(t) = w(i-ct), \; c>0. \label{e13}
\end{eqnarray}
The electric field profile $w(z)$, $z=i-ct$ should satisfy 
\begin{eqnarray}
c\, {dw\over dz}  + {D(w(z)) + v(w(z))\over \nu}\, [w(z-1)-w(z)] 
\nonumber\\
+ {D(w(z))\over \nu}\, [w(z+1)-w(z)] + J - v(w(z)) \leq 0.  
\label{e14}
\end{eqnarray}
We seek a piecewise continuous supersolution which equals a constant, 
$A$, $A\in (E^{(1)}(J),E^{(2)}(J))$, for $z<z_0$, and $w(z) = 
E^{(3)}(J)$ for $z>z_1$, with $z_1>z_0$. For $z_0<z<z_1$, $w(z)$ is an 
unspecified smooth increasing function with $w(z_0) = A$, and $w(z_1) 
= E^{(3)}(J)$. As for subsolutions, we now select conveniently 
the numbers $z_0$, $z_1$, $c$ and $A$ so that (\ref{e14}) holds. Clearly, 
(\ref{e14}) holds for $z+1<z_0$ and for $z-1>z_1$. Suppose that $0<z_1 - 
z_0 < 1$. An analysis of the remaining five possibilities yields the 
following criteria to hold for $w(i-ct)$ to be a supersolution:
\begin{eqnarray}
J-v(A) \leq - {D(A)\over \nu}\, [E^{(3)}(J)-A]\equiv f_3(A;J), 
\label{e15}\\ 
f_4(w;J,A)\equiv -{D(w)+v(w)\over \nu}\, (w-A) \nonumber\\
+ {D(w)\over \nu}\, [E^{(3)}(J)-w] +J-v(w)\leq -\delta,\, \label{e16}\\
\mbox{ for }\quad\quad A\leq w\leq E^{(3)}(J), \nonumber\\
c\, {dw\over dz} \leq \delta .\label{e17}
\end{eqnarray}
Provided such $w(i-ct)$ is found, solutions $E_i(t)$ of (\ref{e7}) with 
$E_i(0)< w(i+\tau)$ will satisfy $E_i(t)< w(i-ct+\tau)$ and propagate 
to the right with speed larger than $c$. $\tau$ is a constant which 
can be conveniently chosen to keep the monopole profile below the 
supersolution. Figure \ref{f9} illustrates the graphical construction of 
the supersolution by checking that (\ref{e15}) and (\ref{e16}) hold for 
particular values of $J$ and $\nu$. 

As in the subsolution case, an upper bound $c^{*}$ for the monopole velocity 
$c$ is estimated by choosing $-\delta(J,A) = $ max$_{A<E<E^{(3)}} f_4(E;J,A)$, 
$z_1-z_0=1$, and $w(z)=[E^{(3)}(J)-A] (z-z_0)$ for $z_0<z<z_1$. Then 
$c^{*} = \delta(J,A)/[E^{(3)}(J)-A]$. In Figure \ref{f4}, $c^{*}$ is 
represented by a line of thick dots for doping $\nu=3$ corresponding to the 
9/4 SL. 

We can now summarize the results obtained from sub and supersolutions; 
see Figures \ref{f8} and \ref{f9}. We find reasonably good upper bounds 
for the absolute value of monopole velocity. Furthermore, for 
$\nu=3$, conditions (\ref{e10}) and (\ref{e11}) hold for $J=0.6$ and  
$A=12$, whereas conditions (\ref{e15}) and (\ref{e16}) hold for $J=0.05$ 
and $A=3$. Therefore, monopoles move downstream for $J \leq 0.05$ and 
they move upstream for $J \geq 0.6$. Direct numerical simulations show 
that: (i) the estimate $J_1 = 0.05$ for the first critical current can 
be improved to $J_1 = 0.08$; and (ii) $J_2 = 0.6$ for the second critical 
current can be improved to $J_2 = 0.54$.

\section{Conclusions and final comments}
\label{sec:dis}
We have presented a theory of monopoles moving downstream or upstream 
on an infinitely long doped, current biased superlattice when the fields
are on the first plateau of the current--voltage characteristic. This 
theory has been corroborated with numerical evidence, which sharpens our
results. Furthermore, we have simulated a 40-well 9nmGaAs/4nmAlAs SL
\cite{kas97} under doping and contact conditions similar to experimental 
ones \cite{bon99}, but under constant current bias conditions. This 
situation is different from the usual case of voltage bias conditions. 
We have obtained that it is possible to observe monopole wavefronts 
moving upstream when the current is kept at large enough levels. Together 
with our theoretical bounds for critical currents and dopings, this 
numerical prediction could be used to set up an experiment to observe 
this striking phenomenon. For this purpose, we would need an initial condition
corresponding to a monopole separating two electric field domains at high 
enough current. In an ideal world, this situation could be obtained by first 
fixing a low dc voltage for the 9/4 sample at a value near the top of one of 
the first branches of the current - voltage characteristics. Then we could 
switch from voltage to current bias conditions. The outcome 
would be a monopole moving upstream until the emitter region is reached. 
Presumably an idea of the field distribution corresponding to this situation 
could be obtained by time-resolved photoluminescence measurements \cite{kas95}.

There are technical problems that must be overcome if
one wants to observe these features in real experiments: 
when we switch, there will always be a Faraday-like inductive pulse which 
will probably perturb the state of the system in an uncontrolled way. 
There are other possible biases we could think of. Under dc voltage bias, 
upstream moving monopoles are probably created for a short time during 
relocation experiment \cite{luo97}. In these experiments, one has a doped
SL with a current-voltage characteristics correponding to multiple stationary 
monopole solution branches. Voltage is set at a particular value near the 
end of a branch, so that the field profile is that of a monopole layer 
connecting a low to a high field domain. Let the monopole layer be located 
at well $i$ (counted from the emitter contact). Then the voltage is suddenly 
and appropriately increased. After a certain time, the field profile settles 
to a new situation corresponding to a monopole layer centered at well $i-1$
\cite{luo97}. This could be an indication of a monopole moving upstream, 
albeit for a short time. To increase this time, we could try to set a hybrid 
bias (between current and voltage bias) by including a finite series resistance 
in our external circuit. Additional theoretical and numerical work is needed 
to explore these possibilities. 

\acknowledgements
We thank A.\ Amann, G.\ Platero and D.\ S\'anchez for fruitful discussions 
and collaboration on the discrete drift-diffusion model. 
LLB thanks S.W.\ Teitsworth for a critical reading of the manuscript and 
helpful comments. This work was supported by 
the Spanish DGES through grant PB98-0142-C04-01, and by the European Union 
TMR contracts ERB FMRX-CT96-0033 and ERB FMRX-CT97-0157.

\appendix
\section{Comparison principle}
\label{theorem}

The main theorem which we use to prove our results in Section \ref{sec:math}
is the following comparison principle: 

\noindent {\bf Theorem A.1}
{\it Let $U_i(t)$ and $L_i(t)$, $i\in Z$, be differentiable sequences such that
\begin{eqnarray}
{d U_i\over dt}-d_1(U_i)[U_{i+1}-U_i]\quad\quad\quad\quad\nonumber\\
-d_2(U_i)[U_{i-1}-U_i] -f(U_i)\geq \nonumber \\
{d L_i\over dt}-d_1(L_i)[L_{i+1}-L_i]\quad\quad\quad\quad\nonumber\\
-d_2(L_i)[L_{i-1}-L_i] -f(L_i), \label{l} \\
U_i(0) > L_i(0). \nonumber
\end{eqnarray}
where $f$, $d_1>0$ and $d_2>0$ are Lipschitz continuous functions. Then,
$$ U_i(t) > L_i(t),\quad t>0, i\in Z $$}

In our discrete drift-diffusion model,
\begin{eqnarray}
 d_1(E) = {D(E)\over\nu}\,,\, d_2(E) = {D(E)+v(E)\over\nu}\,,\nonumber\\
f(E) = J-v(E).\nonumber 
\end{eqnarray}

{\bf Proof:} The proof is by contradiction. Set $W_i(t)=U_i(t)-L_i(t)$.
At  $t=0$, $W_i(0)>0$ for all $i$. Let us assume that $W_i$ changes sign 
after a certain  minimum time $t_1>0$, at some value of $i$, $i=k$.
Thus $W_k(t_1)=0$ and $dW_k/dt\leq 0$, as $t\to t_1$. We shall show that 
this is contradictory. At $t=t_1$, there must be an index $m$ (equal or 
different from $k$) such that $W_m(t_1)=0$, while its next neighbor 
$W_{m+j}(t_1)>0$ ($j$ is either 1 or -1), and $W_i(t_1)=0$ for all 
indices between $k$ and $m$. For otherwise $W_k$ should be identically 
0 for all $k$. Equation (\ref{l}) implies
\begin{eqnarray}
{d W_{m}\over dt}(t_1) \geq d_1(U_{m}(t_1))\, W_{m+1}(t_1)
\quad\quad\nonumber\\
+ d_2(U_{m}(t_1))W_{m-1}(t_1) >0. \nonumber 
\end{eqnarray}
This contradicts the fact that $dW_m/dt$ should have been nonpositive as 
$t\to t_1$, for $W_m(t_1)$ to have become zero in the first place.  \\

\noindent {\bf Corollary A.1}
{\it Any solution $E_i(t)$ of (\ref{e7}) with initial
data $E_i(0)\in (E^{(1)}(J),E^{(3)}(J))$ satisfies $E_i(t)\in (E^{(1)}(J),
E^{(3)}(J))$ for $t>0$.}

{\bf Proof:} Apply Theorem A.1 first with $L_i=E^{(1)}(J)$ and
$U_i=E_i$, then with $L_i=E_i$ and $U_i=E^{(3)}(J)$.\\

\noindent {\bf Corollary A.2}
{\it If $E_i(0)$ is monotone increasing, that is,
$E_i(0)< E_{i+1}(0)$, then, $E_i(t)$ is also monotone
increasing, i.e., $E_i(t)< E_{i+1}(t)$ for $t>0$.}

{\bf Proof:} Apply Theorem A.1 with $L_i=E_i(t)$ and
$U_i=E_{i+1}(t)$.\\

\noindent {\bf Remark}. Strict inequalities in these theorems can be replaced
by inequalities and the corresponding statements still hold. However the 
proofs become rather more technical and involved.

\section{Bounds for critical doping values}
\label{nuc}
We want to estimate the curves $\nu_{1b}(J)$ and $\nu_{2b}^{\pm}(J)$ 
defined in Section \ref{sec:math}. To estimate $\nu_{2b}^{+}(J)$, assume 
that $J\to 1-$ and $\nu$ is large. The left tail of a monopole is 
pinned if Eq.\ (\ref{c-}) holds. For large currents, (\ref{c-}) certainly 
holds if the curve corresponding to the left side of the 
inequality is above that of the right hand side, for $E=1$ (this is possible
because $D(E)$ decreases rapidly to zero as the field increases). Setting
$E=1$ in (\ref{c-}), we obtain 
$$
{D(1)\, [1-E^{(3)}(J)]\over \nu} > J-1.
$$
In turn, this implies $\nu>\nu_{2b}^{+}(J)$, defined in (\ref{nu2}). 
This argument fails for the small values of $J$ used to draw Figure 
\ref{f7}. We believe that quite different reasoning is needed to 
estimate $\nu_{2b}^{-}(J)$. 

\begin{figure}
\begin{center}
%\centerline{\hbox{\psfig{figure=jcritlog.ps,width=7.0cm}}}
\fbox{
\epsfig{file=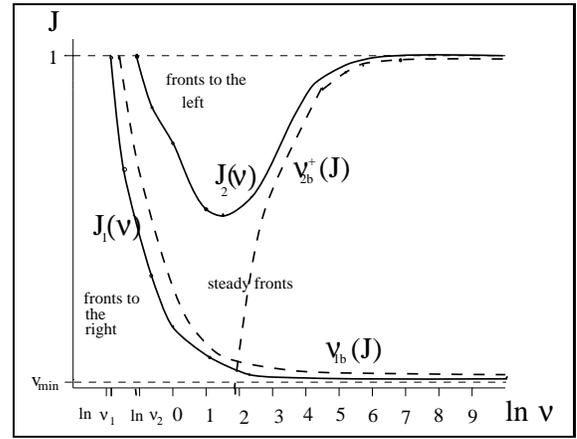,width=7cm}
}
\vspace{0.5 cm}
%\centerline{\hbox{\psfig{figure=fig10c.ps,width=7.0cm}}}
\caption{ Critical currents $J_1$ and $J_2$ as functions of the 
dimensionless doping $\nu$. Monopoles move downstream for $v_m<J<J_1(\nu)$, 
are stationary for $J_1(\nu)<J<J_2(\nu)$, and move upstream for $J_2(\nu)
<J<1$. Dashed lines in this figure represent the bounds $\nu_{1b}(J)$ and
$\nu_{2b}^{+}(J)$.}
\label{f3}
\end{center}
\end{figure}

\begin{figure}
\begin{center}
\fbox{
\epsfig{file=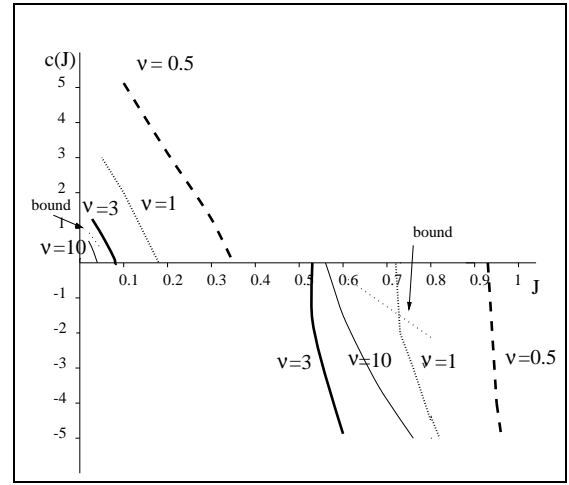,width=7cm}}
\vspace{0.5 cm}
\caption{Velocity of a monopole wavefront as a function of $J$ for four 
doping values $\nu=0.5$, 1, 3, and 10. Monopoles with negative velocity move
upstream. For doping $\nu=3$ corresponding to the 9/4 SL, we have also 
represented bounds for the velocity as lines of thick dots. }
\label{f4}
\end{center}
\end{figure}

\begin{figure}
\begin{center}
\fbox{
\epsfig{figure=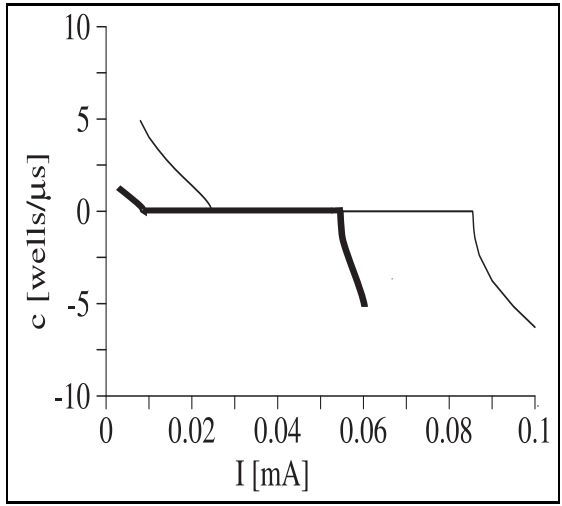,width=7.0cm}}
\vspace{0.5 cm}
\caption{Monopole velocity as a function of current for the 9/4 SL. 
Comparison of results for the discrete drift-diffusion model (1) - (2) 
with $\nu=3$ (thick line), and those obtained by using the exact tunneling 
current $ eJ(n_i,n_{i+1},F_i)$ (thin line) instead of approximation (3). }
\label{f5}
\end{center}
\end{figure}

\begin{figure}
\begin{center}
\fbox{
\epsfig{file=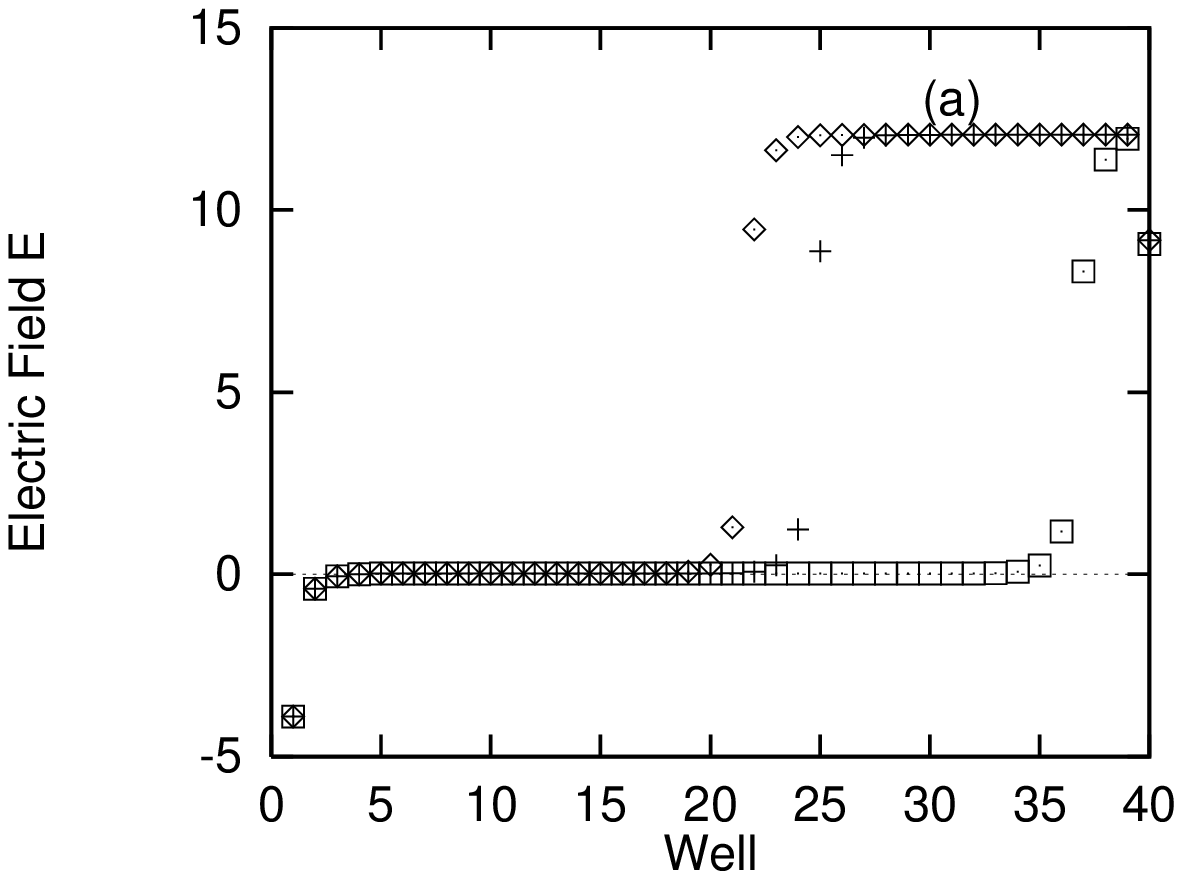,width=7cm}}
\vspace{0.5 cm}
\fbox{
\epsfig{file=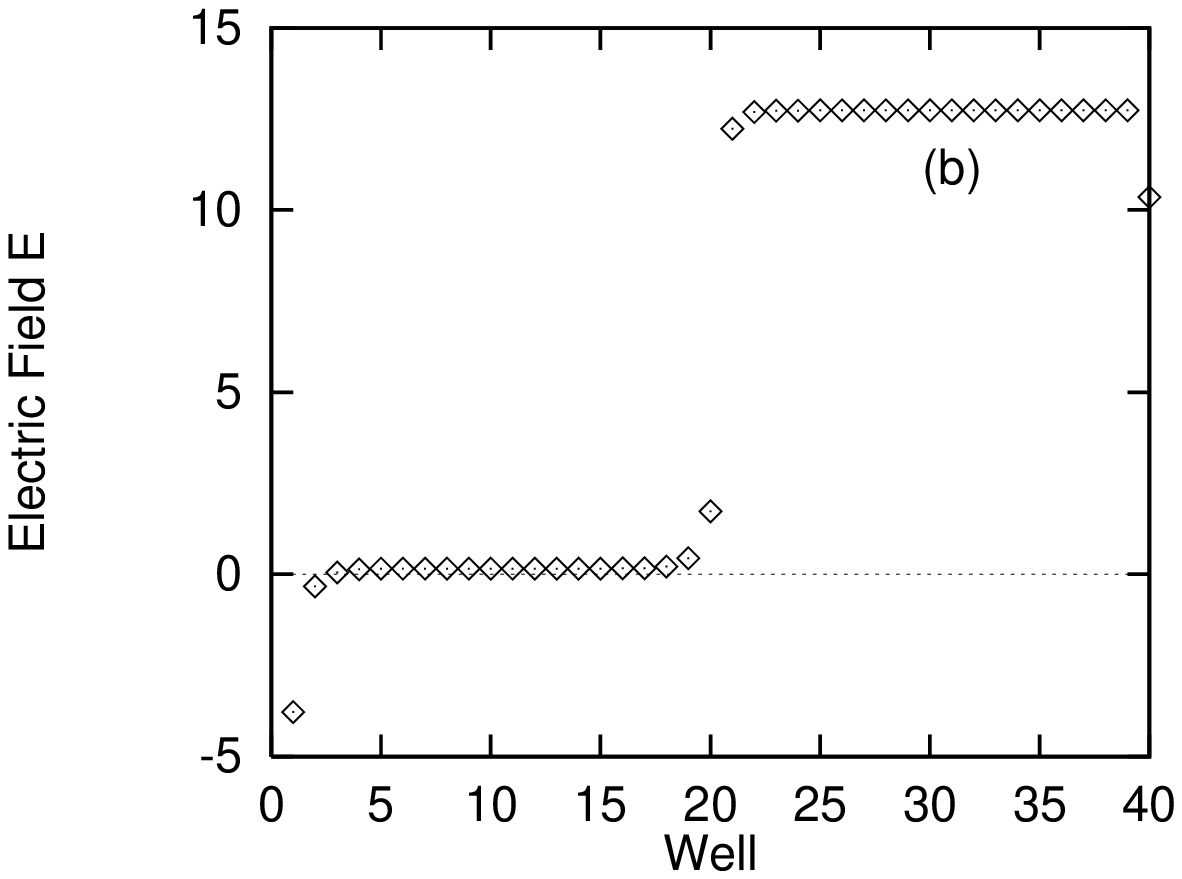,width=7cm}}
\vspace{0.5 cm}
\fbox{
\epsfig{file=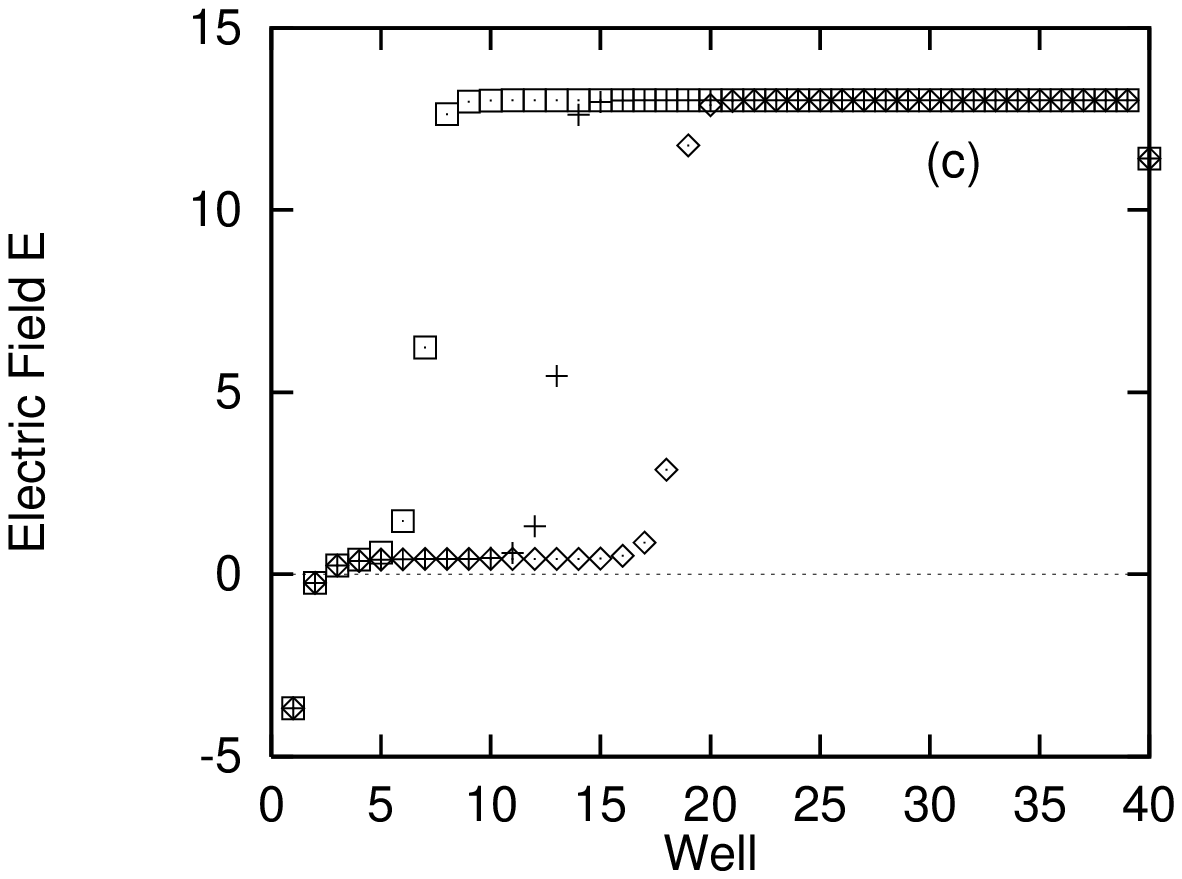,width=7cm}}
\vspace{0.5 cm}
\caption{ Numerical simulations of the drift-diffusion model with 
realistic boundary conditions at the contact regions. (a) Monopole 
moving downstream for $J=0.023$, (b) stationary monopole for $J=0.3$, and 
(c) monopole moving upstream for $J=0.9$. In all cases, diamonds correspond 
to the profile at $t=0$ and squares to the profile at the largest positive 
time.}
\label{f6}
\end{center}
\end{figure}

\begin{figure}
\begin{center}
\fbox{
\epsfig{file=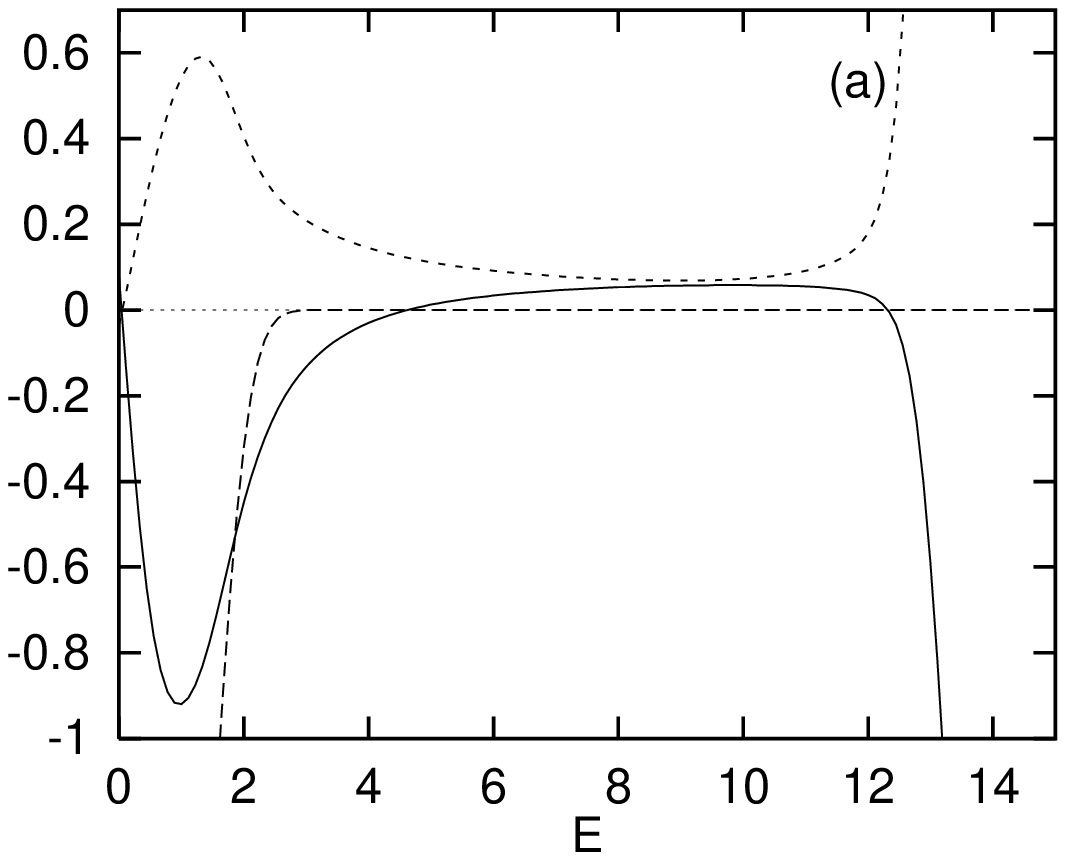,width=7cm}}
\vspace{0.5 cm}
\fbox{
\epsfig{file=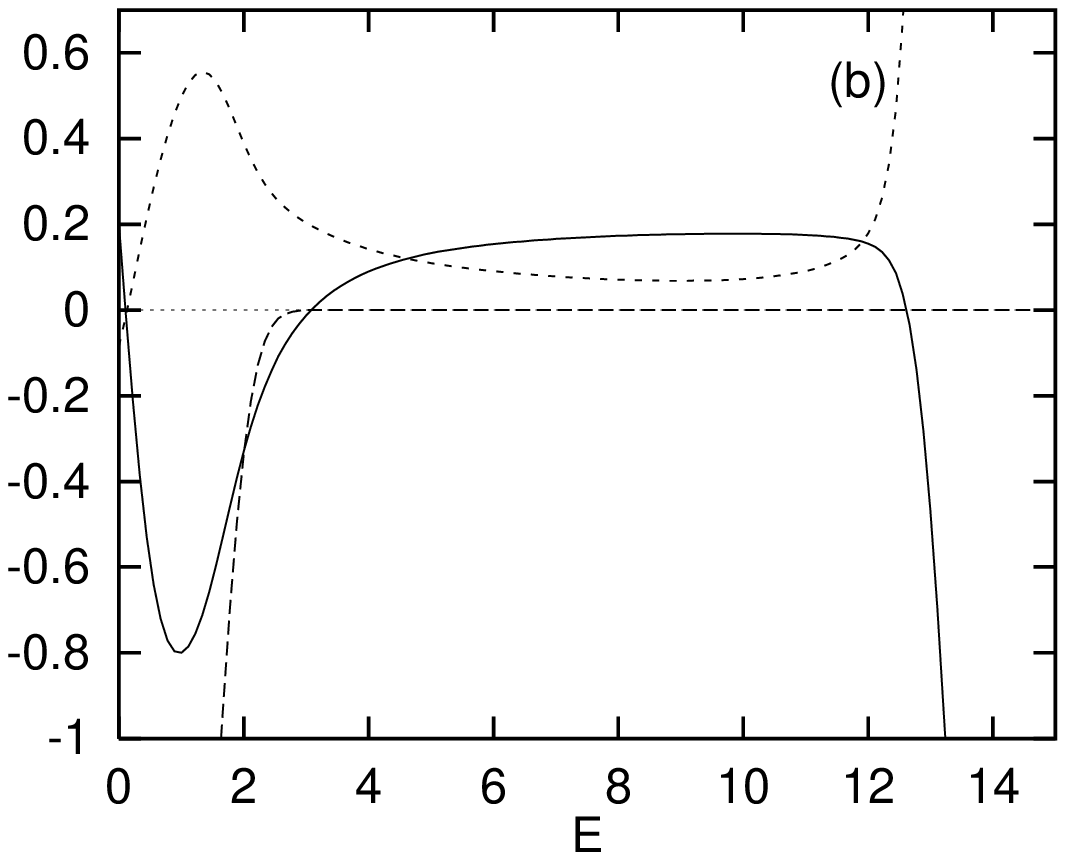,width=7cm}}
\vspace{0.5 cm}
\fbox{
\epsfig{file=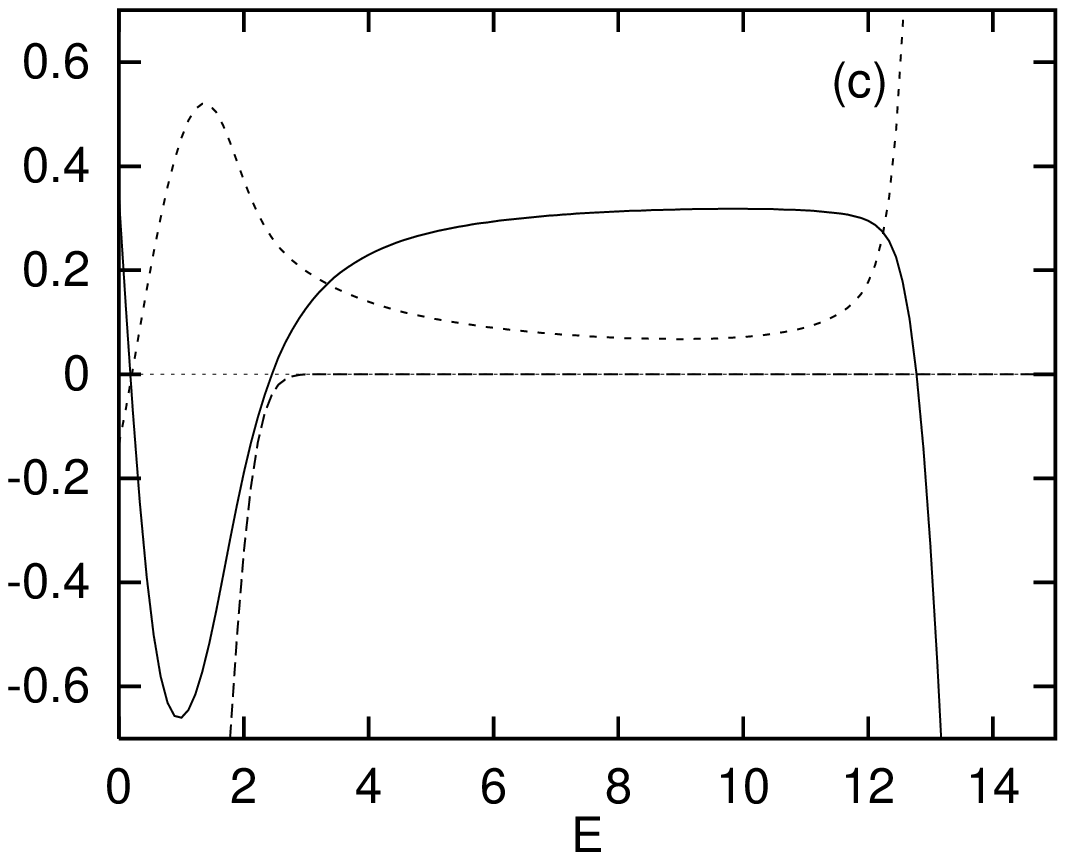,width=7cm}}
\vspace{0.5 cm}
%\hbox{\psfig{figure=fig10b.ps,width=7.0cm}}}
%\centerline{\hbox{\psfig{figure=fig10c.ps,width=7.0cm}}}
\caption{(a) Functions of the electric field establishing pinning of 
the monopole tails for $\nu=3$, $J=0.08$. Solid line: $J-v(E)$, dashed line: 
$D(E)\, [E - E^{(3)}(J)]/\nu$, dotted line: $[D(E)+v(E)]\, [E - 
E^{(1)}(J)]/\nu$. (b) Same plots as in (a) for $\nu =3$, $J=0.2$.
(c) Same plots as in (a) for $\nu =3$, $J=0.34$. }
\label{f7}
\end{center}
\end{figure}

\begin{figure}
\begin{center}
\fbox{
\epsfig{file=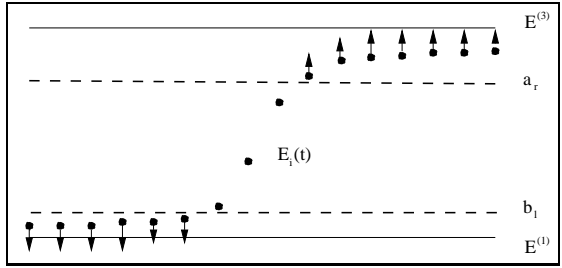,width=7cm}}
\vspace{0.5 cm}
%\hbox{\psfig{figure=fig10b.ps,width=7.0cm}}}
%\centerline{\hbox{\psfig{figure=fig10c.ps,width=7.0cm}}}
\caption{A monopole field profile. The field values $b_l$ and 
$a_r$ are indicated by horizontal dashed lines. Vertical arrows 
indicate that the field at those SL periods either (i) decrease toward
$E^{(1)}(J)$, and therefore the monopole left tail is pinned; or (ii)
increase toward $E^{(3)}(J)$, and therefore the monopole right tail 
is pinned. }
\label{fpin}
\end{center}
\end{figure}

\begin{figure}
\begin{center}
\fbox{
\epsfig{file=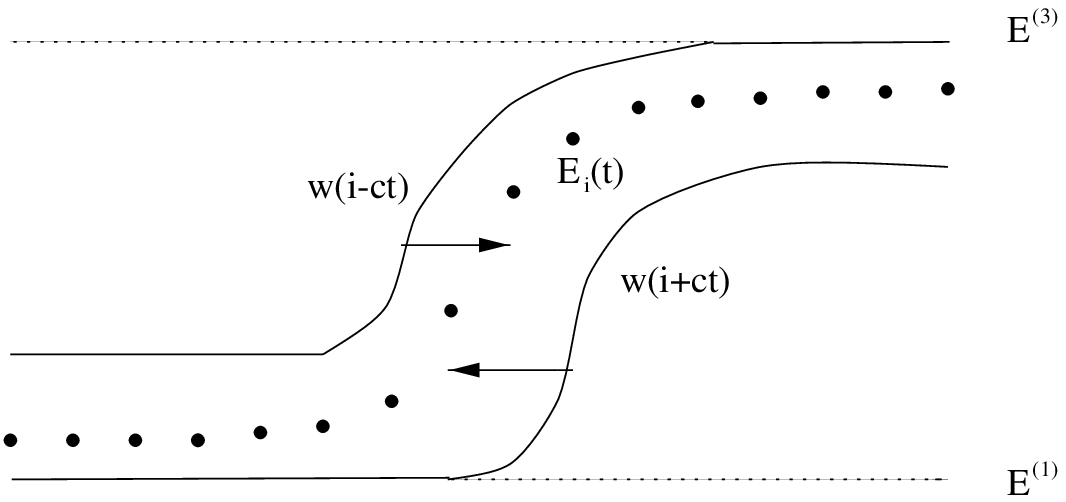,width=7cm}}
\vspace{0.5 cm}
\caption{A monopole field profile, sub and supersolutions. The supersolution 
is always above the real values of the field, and therefore it pushes the
monopole to the right. The subsolution pushes (from below) the
monopole to the left. }
\label{fsbsp}
\end{center}
\end{figure}

\begin{figure}
\begin{center}
\fbox{
\epsfig{file=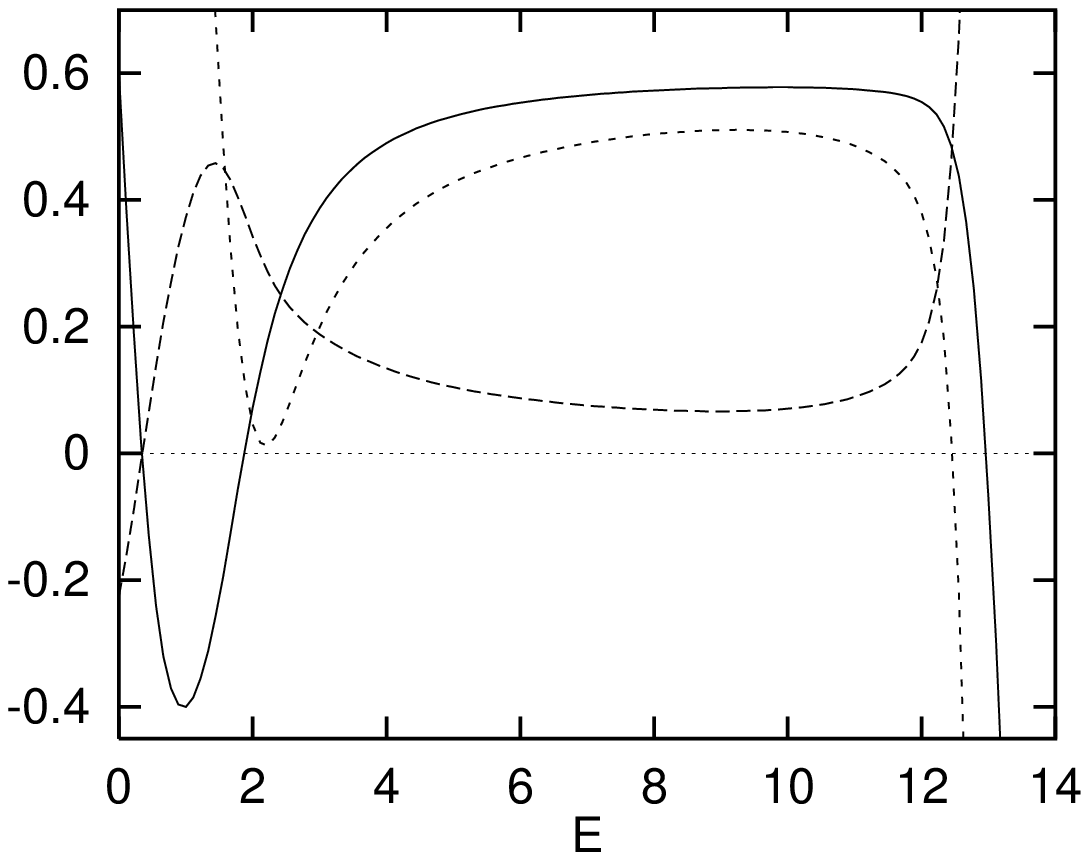,width=7cm}}
\vspace{0.5 cm}
\caption{Curves determining the subsolution for $\nu =3$, $J=0.6$, 
$A=12$.
Solid line: $J-v(E)$; dashed line: $f_{1}(E;J)$; dotted line:   
$f_2(E;J,A)$. }
\label{f8}
\end{center}
\end{figure}

\begin{figure}
\begin{center}
\fbox{
\epsfig{file=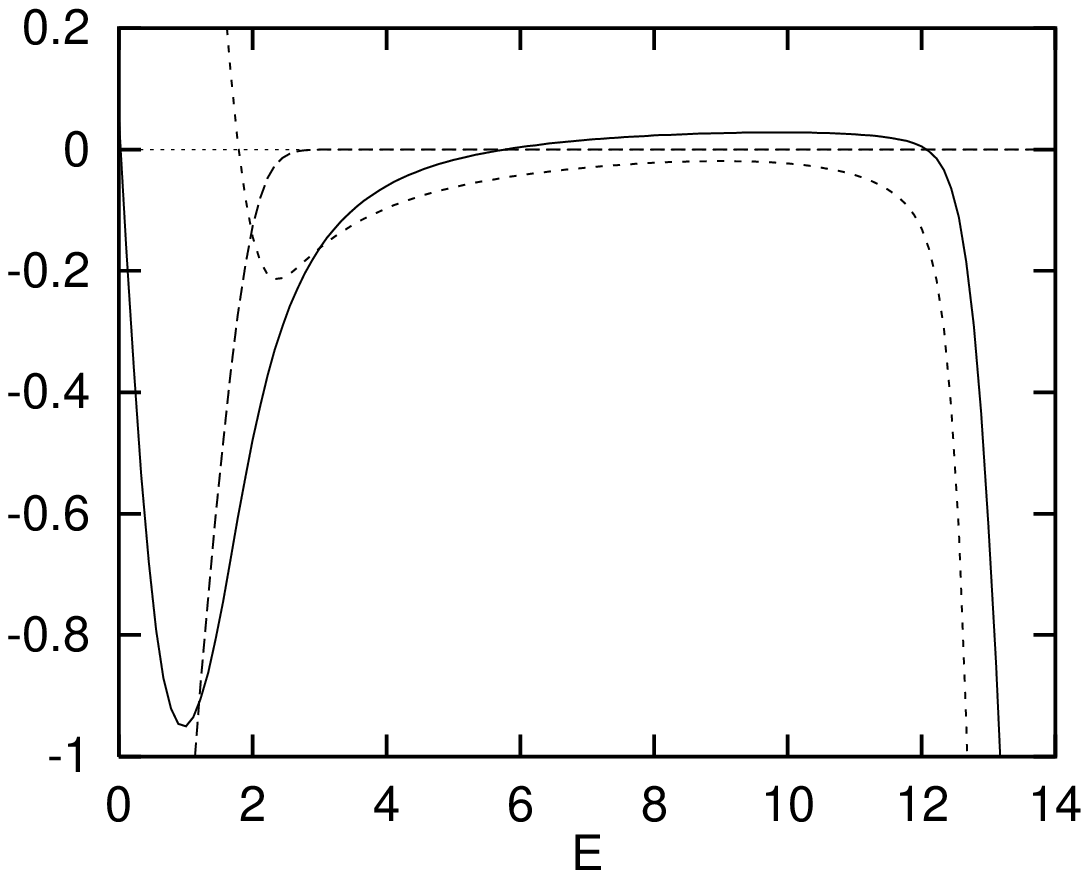,width=7cm}}
\vspace{0.5 cm}
\caption{Curves determining the supersolution for $\nu =3$, $J=0.05$, 
$A=3$.
Solid line: $J-v(E)$; dashed line: $f_{3}(E;J)$; dotted line:  
$f_4(E;J,A)$. }
\label{f9}
\end{center}
\end{figure}

The same argument yields our estimate $\nu_{1b}(J)$ of (\ref{nu1}). For 
(\ref{c+}) to hold, the curve corresponding to the left side of the 
inequality should be below that of the right hand side for $E=E_m$. 
As $D(E_m)\approx 0$, we obtain
$$ 
{v(E_{m})\over\nu}\, [E_m - E^{(1)}(J)] < J - v(E_m),
$$
which yields (\ref{nu1}). Fig.\ \ref{f3} shows that the bound (\ref{nu1}) 
is reasonably good for all eligible values of $\nu$ and $J$.

%\newpage

%\newpage

%\newpage

\end{multicols}	
\end{document}